# Workforce Development Through Research-Based, Plasma-Focused Activities

E G Kostadinova [1], Shannon Greco [2], Maajida Murdock [3], Ernesto Barraza-Valdez [4], Hannah R Hasson [5], Imani Z West-Abdallah [5], Cheryl A Harper [6], Katrina Brown [7], Earl Scime [8], Franklin Dollar [4], Carl Greninger [9], Bryan Stanley [10], Elizabeth Oxford [11], David Schaffner [12], Laura Provenzani [13], Chandra Breanne Curry [14], Claudia Fracchiolla [15], Shams El-Adawy , Saikat Chakraborty Thakur [1], Dmitri Orlov [17], Caroline Anderson [18]

[1] Auburn University, AL, USA; [2] Princeton Plasma Physics Laboratory; [3] Morgan State University; [4] University of California, Irvine; [5] University of Rochester; [6] Greensburg Salem High School; [7] University of Pittsburgh at Greensburg; [8] West Virginia University; [9] Microsoft; [10] Michigan State University; [11] University of Michigan; [12] Bryn Mawr College; [13] University of Alabama in Huntsville; [14] SLAC National Accelerator Laboratory; [15] American Physical Society; [16] Kansas State University; [17] University of California, San Diego; [18] Fusion Industry Association

Contact e-mails: egk0033@auburn.edu, sgreco@pppl.gov

**Abstract.** This report is a summary of the mini-conference Workforce Development Through Research-Based, Plasma-Focused Science Education and Public Engagement held during the 2022 American Physical Society Division of Plasma Physics (APS DPP) annual meeting. The motivation for organizing this mini-conference originates from recent studies and community-based reports highlighting important issues with the current state of the plasma workforce. Here we summarize the main findings presented in the two speaker sessions of the mini-conference, the challenges and recommendations identified in the discussion sessions, and the results from a post-conference survey. We further provide information on initiatives and studies presented at the mini-conference, along with references to further resources.

PREFACE

Here we present the summary of the mini-conference *Workforce Development Through Research-Based, Plasma-Focused Science Education and Public Engagement* held during the American Physical Society Division of Plasma Physics (APS DPP) conference in Spokane, WA, from Oct. 17 to Oct. 21, 2022. The main goal of the mini-conference was to initiate communication among a diverse selection of colleagues interested in promoting knowledge about plasma (the ionized state of matter), educating and training the plasma workforce, and improving the health of the plasma community. The title of this report was revised to *Workforce Development Through Research-Based, Plasma-Focused Activities* to recognize that the mini-conference participants shared expertise beyond the original focus on *Science Education and Public Engagement*.

Throughout this report, we use general words like "plasma STEM" and "plasma professionals" instead of the more common "plasma science and fusion energy" and "plasma physicists" to emphasize the expanding role of plasma throughout all STEM fields and the increasing role of physics education researchers, science communicators, K-12 educators, and other colleagues involved in broader impact activities within the plasma community. We further recognize that the word "outreach", commonly used as an umbrella term for the activities discussed here, implies one-way communication. This document promotes two-way interactive methods to bring science "alive", spark interest, and stimulate curiosity, which can be better described by the word "engagement". Finally, the authors recognize that this report reflects the opinions of the mini-conference participants and colleagues who engaged in related discussions. Thus, the needs and recommendations outlined below are not meant to be an exhaustive list but the basis for further discussions and motivation for actions within the community.

Section I provides a brief motivation for organizing this mini-conference along with some technical information. In Sec II, we summarize the main findings presented in the two speaker sessions of the mini-conference, the challenges and recommendations identified in the discussion sessions, and the results from a post-conference survey. Section III discusses strategies and current activities in more detail, including examples and further resources. The Appendix includes the mini-conference program and information on data availability.

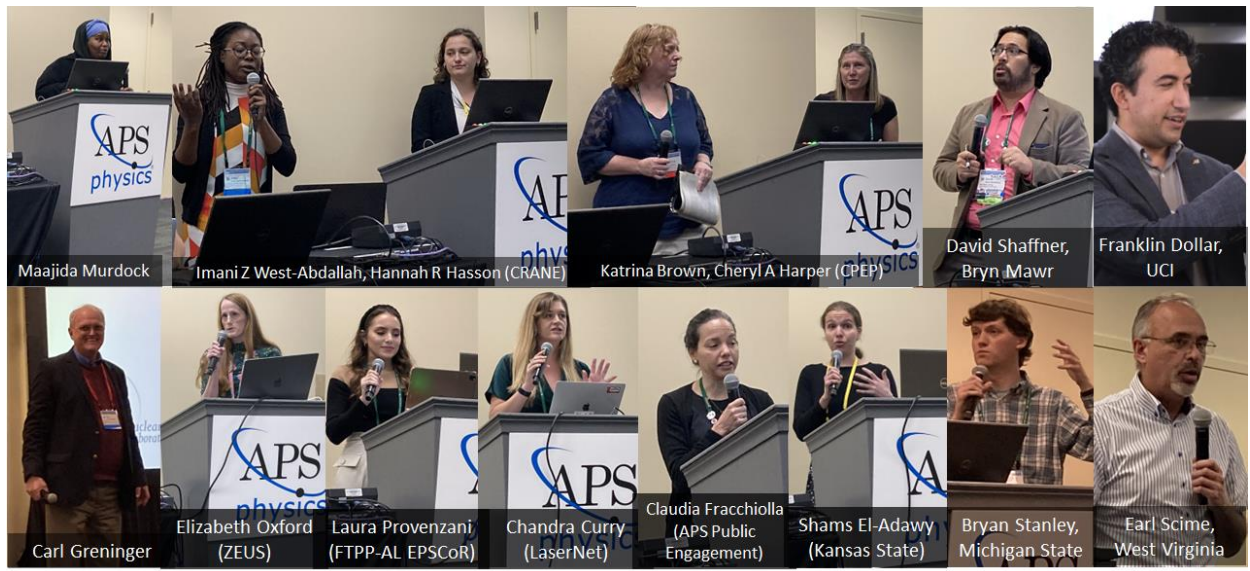

Fig. 1 Speakers at the mini-conference *Workforce Development Through Research-Based, Plasma-Focused Science Education and Public Engagement* held at the 2022 APS DPP.

The mini-conference was conducted in a hybrid format, including in-person and remote participation. All talks were broadcast live and recorded. The presentation slides are available in a shared Google drive to all participants in our Slack channel [1]. This report is submitted for publication in the special issue of Physics of Plasmas dedicated to the 2022 APS DPP conference. The document is also available on ArXiv.

---

[1] Use this link to join the Slack channel: https://join.slack.com/t/plasmanet/shared_invite/zt-1oddah56s-FjpPkwAXq99SkxqbetZCEw

## I. INTRODUCTION

In the past two decades, two assessments on the workforce needs for plasma science and fusion energy have identified major issues with the declining number of plasma faculty, the small number of departments and institutions teaching plasma, and the slow production rate of qualified personnel [1], [2]. Both reports recognized the importance of outreach programs at all educational levels, starting with pre-college education and public engagement. The recent National Academies of Science decadal assessment of plasma science found that plasma-specific educational and research programs that also provide opportunities to diverse and less advantaged populations are needed to ensure a critically populated plasma science and engineering workforce [3]. The recent Fusion Energy Sciences Advisory Committee report Powering the Future: Fusion & Plasmas recommended that policy changes be developed and implemented to allow for "discipline-specific workforce development" [4]. In response to these findings, the non-profit organization Coalition for Plasma Science (CPS) and the Princeton Plasma Physics Lab (PPPL) co-sponsored an APS DPP mini-conference focused on addressing the workforce needs of the plasma community.

## II. FINDINGS, CHALLENGES, RECOMMENDATIONS, & RESPONSIBILITIES

This mini-conference aimed to establish best practices for science education and public engagement which can be geared toward the workforce needs of the plasma community. In other words, while various successful STEM programs exist, there is a need to incorporate them into plasma-focused efforts that are well-coordinated and endorsed by the plasma community. The first half-day of the mini-conference talks discussed the crucial role of informal science education in K-12, engaging and retaining marginalized communities, providing professional training in high school, and building successful outreach programs in economically depressed areas. In the second half-day of the mini-conference, talks were focused on adapting public engagement strategies to the needs of various entities, including large user facilities, a networks of user facilities, small colleges, and an EPSCoR state-wide collaborations. This section summarizes findings, challenges, and recommendations related to these topics. In addition, a suggested breakdown of responsibilities is included after each recommendation.

### A. Informal Plasma Education, Engagement, and Alternative Education

The primary purpose of public engagement and training, encompassing informal learning experiences, "outreach," internships, etc., is typically seen as attracting and developing new audiences to support the growth of the field. Ignoring historically excluded communities from these efforts is not only morally wrong, but it hinders the development of the field. STEM disciplines have historically excluded and, at times, outright abused certain communities [5]. Without including these communities in conversations about science, talented individuals choose other careers, mistrust builds, public support of research funding dwindles, science is ignored in policy decisions, and ultimately, society suffers. A major goal for conducting the mini-conference was to initiate building a network, or an ecosystem, where we hope to share resources and experiences that would help us be more effective and expand access to the array of opportunities in plasma STEM. We hope that frequent communication will result in pushing each other to be intentional about how and with whom we engage.

1. Informal Plasma Education Activities (IPEA) and Plasma Learning Ecosystems (IPEs)

Informal science education activities (ISEA) have played an essential role in creating and recruiting the next generation of scientists [6]. While K-12 teachers are constrained by the curriculum, textbooks, and test requirements at the national and state level, guest researchers have the freedom to promote creative and culturally relevant learning experiences of science. Researchers visiting the classroom can improve student/teacher engagement by explaining concepts that may be new and/or missing from the K-12 curriculum and by communicating exciting results from the latest research work, making science engaging and contemporary.

Studies of educators' conceptualizations of STEM education have revealed commonalities along the need for: (i) interdisciplinary connections, (ii) new, ambitious instructional practices, (iii) and engagement of students in real-world problem solving [7], [8]. Some of the greatest strengths of plasma research – its highly interdisciplinary nature, connection to industry, rapid growth, and significant impact on some of the latest technologies – can directly address some of these needs. However, due to the breadth of plasma topics (ranging from space and astrophysics to fusion energy, to low-temperature plasma technology, and more), it is often challenging to communicate consistent information to K-12 teachers/students and the public. This issue is not unique to plasma or STEM, but also a general question of strategy research, where construct clarity has been identified as foundational to knowledge accumulation [9]. The NAS study titled *Framework for K–12 Science Education* [10], which became the first step in establishing the Next Generation Science Standards (NGSS) [11], emphasized that science education in the US should be organized systematically across multiple years of school, emphasize depth of concepts, and provide students with engaging opportunities to experience how real-life science is done.

Based on these findings, we argue that to improve consistency in building plasma-relevant knowledge in K-12 education, the plasma community should work towards establishing a minimal level of consensus when introducing terminology in both formal and informal settings. Maintaining consistency enables the audience to connect and engage with ideas when the same topics are discussed on different occasions. Consistent terminology should further be supported with relatable examples and appropriate context to allow flexibility of communication. However, care should be taken to periodically update terminology as scientific discourse and ideas evolve. Terminology, examples, and context cannot be effective without developing a relationship with the community. These should be co-created with input from teachers, students and community leaders to avoid a siloed perspective that maintains the separation between "science" and the "public."

One way to address the above challenges is to study best practices in ISEA [6], [12]–[14]. Two examples of successful ISEA activities include the proliferation of after-school informal science programs and the establishment of STEM Learning Ecosystem (SLE) communities. Both types of activities are founded in cross-sector partnerships between schools, after-school and summer programs, libraries, museums, and businesses, among others. It has been demonstrated that after-school informal science programs support and promote positive attitudes about science among students from both traditional and underrepresented groups [15], [16]. It has also been shown that university student participation in ISEA increases the likelihood of retention in a career related to the field [17]–[20]. Finally, a survey among 37 SLEs shows that most communities report a willingness to create a shared vision around data collection, which will help both researchers and educators track, understand, and improve the quality of STEM education and its impact [21]. Based on these examples, we propose supporting the proliferation of plasma-focused after-school programs, establishing Plasma Learning Ecosystem (PLE) communities, and creating an open-source repository for Informal Plasma Education Activities (IPEA) available to the community.

At least four stakeholders would benefit from these efforts – the audiences, the informal educators, the institutions, and the plasma community. The formation of the PLEs and IPEA repository would allow informal plasma educators to share experiences with diverse audiences, standardize terminology, develop impact assessment tools, and send coordinated messages to the public. This will benefit the proliferation of plasma-focused after-school programs at K-12 institutions, which will offer an engaging plasma education experience complimentary to the existing curriculum. Other stakeholder institutions, such as universities, national labs, and industries, will further benefit from the shared IPEA repository as it will provide much-needed support for newly-employed informal plasma educators. Finally, an increased role of well-coordinated, plasma-specific informal education could impact the public opinion of plasma, help increase government funding for plasma, and expand the plasma workforce.

Effective engagement of all audiences mentioned above is inseparable from dedicated strategies for Diversity, Equity, Inclusion, Accessibility, and Social Justice (DEIA SJ). The growth of the nuclear fusion and plasma technology industries in the past few years has shown a pressing need for broadening participation in this developing workforce [3]. However, the lack of diversity within the plasma community has been detrimental to encouraging historically underrepresented, marginalized, first-generation students, and students of color to join the plasma STEM fields [1], [3], [22]. This, in turn, depletes the pathways from K-12 and higher education to the breadth of employment opportunities in industries, national labs, government, and academia. The addition of global cataclysms, such as the COVID-19 pandemic, lead to long-term effects (medical, economic, and educational) that disproportionately affect marginalized communities [23], which creates an even larger barrier to joining the plasma community.

The different audiences (including K-12 students and teachers, the public, the press, and the government) will require highly flexible and adaptive plasma education strategies. Those strategies should take into consideration an understanding of the audience's existing knowledge and needs, determining the best medium to engage with the audience, including appropriate venues, online platforms, and a curriculum/content that provides relatable examples, connections to the audiences everyday life, career prospectives, and engaging on a personal level.

**Challenges, Recommendations and Responsibilities**

*Challenge:* The current K-12 public education curriculum does not offer consistent and well-coordinated learning experiences of plasma STEM, which severely impedes the ability to engage students from diverse cultural and socioeconomic backgrounds.

*Recommendation:* Following the example of ISEA, the plasma community would benefit from establishing Plasma Learning Ecosystems (PLEs) and creating an open-source repository for Informal Plasma Education Activities (IPEA) available to the community.

*Responsibility:* Government funding agencies (e.g., NSF, DOE) should sponsor the establishment and maintenance of the open-access IPEA repository. PLEs should be established with resources from large plasma collaborations (e.g., user/collaborative facilities, centers, Track II EPSCoR grants, etc.) Effective coordination across stakeholders should be maintained by plasma community groups (e.g., APS DPP Education and Outreach, Coalition for Plasma Science, Fusion University Association, etc.)

## 2. Plasma Training Programs

In addition to traditional informal education activities, such as presenting talks and demos in K-12 science classes, organizing museum visits and field trips, and organizing science cafés, the community can benefit from the proliferation of alternative education projects. A remarkable example is the high-school fusion program established and maintained by Carl Greninger in Seattle, WA [24], [25]. For over twelve years, Mr. Greninger has pioneered a private nonprofit helping high school students win over $800,000.00 in scholarships through his non-traditional after-school educational program. The students gain hands-on laboratory experience by growing a Farnsworth fusor (Fig. 2) into a cutting-edge fusion research device. The device is then leveraged for oncology, material science, and plasma physics investigations. Mr. Greninger's program is *local* and *fully self-funded*, showing the important impact private citizen donors can have on the workforce development for plasma.

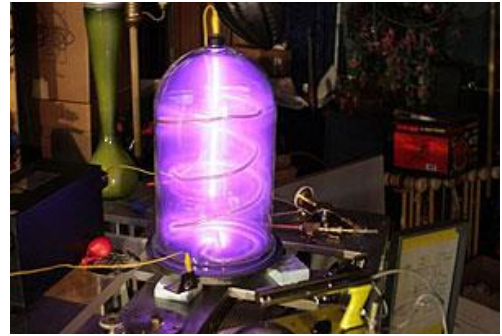

Fig. 2 The fusor machine built by Mr. Greninger and his students.

Based on this example, we propose that strategies should be developed to increase engagement of private citizens interested in improving society through plasma STEM. Private citizens can be involved as volunteers, donors, and/or advocates for plasma-focused programs in their local community. The impact from such programs can be maximized through collaborations with local universities, national labs, and businesses, to develop content and training strategies that will expand academic and employment options for the students. Fundraising efforts and student engagement can be led by plasma science education coordinators, local chapters of professional plasma communities, and plasma nonprofit organizations. Care should be taken to expand access to such programs to students from underrepresented groups and lower socioeconomic status. Partnerships between private donors, local businesses, and Historically Black Colleges and Universities (HBCUs) and Minority Serving Institutions (MSIs) within the same state can help developing plasma training programs for students from local high-schools.

---

**Challenges, Recommendations and Responsibilities**

*Challenge:* Currently, there is no straightforward formal way to solicit and receive private funds for plasma-focused informal/alternative education, severely limiting who has access to such experiences.

*Recommendation:* The plasma workforce will benefit from the establishment and proliferation of local, privately-funded after-school programs focused on hands-on plasma projects. Such programs should target students from underrepresented groups and lower socioeconomic status.

*Responsibility:* Private citizens should be engaged as sponsors, volunteers, and advocates for such programs. Science education coordinators, local chapters of professional communities, and nonprofit organizations should manage fundraising efforts and student engagement. Partners from academic institutions (including HBCUs and MSIs), national labs, and local businesses should help develop content/training for the programs.

---

Providing culturally relevant plasma education experience is a challenge, especially in areas where the majority of student demographics typically come from lower socio-economic status. Student engagement and retention in such areas requires establishing long-term, sustainable, hands-on programs through schools, scouting organizations, 4-H, and direct parental engagement. The establishment and proliferation of such programs focused on plasma-relevant activities can learn from similar programs in adjacent STEM fields. An outstanding example is the FIRST Robotics program at the University of West Virginia [26], which has demonstrated remarkable success in engaging students from low-income families in engineering-focused activities (e.g., via participation in robotics competitions) and motivating them to pursue higher education. The basics of electronics, programming, and problem solving can be effectively taught in similar programs focused on plasma applications.

It has been shown that the performance metrics of marginalized groups strongly depend on preparatory levels even before joining the programs (which could be due to a vast array of socio-economic conditions that preferentially affect marginalized communities) and not due to the skillset of the individuals [27]. Numerous broadening participation programs are highly focused on training specific skills that may not otherwise be widely accessible to students from marginalized groups. However, training without building a welcoming community is not a fruitful strategy for engagement and retention.

In addition to training technical skills, programs aiming to broaden participation in plasma should implement clear strategies for establishing community and relevance to the students. Such strategies include near-peer and peer mentorship, culturally relevant research experiences, and workshops enabling personal growth and wellness [28]. Organizations within the plasma community, such as The Computational Research Access Network (CRANE) [29], are outstanding examples of these principles. CRANE is a group of graduate students, postdocs, and professionals from marginalized communities who combine training with mentorship, and representation in a welcoming environment. In their talk at the mini-conference [30], CRANE members emphasized the importance of representation and role models in the leadership of their organization.

Another great example of a community-led organization focused on broadening participation is the Physics & Astronomy Community Excellence (PACE) [31] at UC Irvine's Department of Physics and Astronomy. PACE is a graduate student-led mentoring organization - peer-mentoring programs by students for students. Their programs include 1-on-1 grad peer mentoring, workshops for incoming graduate students, undergrad-grad mentoring centered on applying to graduate programs), and undergrad-undergrad peer mentoring. PACE is co-sponsored by the Department of Physics and Astronomy the National Osterbrock Leadership Program (NOLP) [32] showing the importance of both local and national support.

Such efforts within the plasma community can also be government funded through career awards, Broader Impact activities [33] on NSF grants and the recently established DOE PIER plans [34]. Government funding through these channels is much appreciated. However, it is important to note here that, while inseparable, public engagement and training efforts do not represent the entirety of DEIA SJ work. There is much work to be done to address systemic racism, sexism and genderism, improving hiring and promotion practices, promoting diversity in leadership positions, and acknowledging the painful role of slavery in the history of US institutions. Public engagement and education is one piece of these efforts and should not consume more than its share of funds and other resources. In other words, participation in outreach activities should be necessary, but not sufficient to fulfil the criteria of broader impact or PER plans. In addition, these plans should include dedicated efforts to address systematic issues within the sponsored institutions.

A main challenge for many of these programs is sustainability through consistent finding sources and continuous commitment from higher education institutions. Another issue arises from lack of recognition for informal education efforts during hiring, performance assessments, and promotions of physicists. Finally, while ground-up initiatives and representative leadership are essential for the engagement and retention of marginalized groups, established and empowered leaders within the community should share the responsibility through measurable actions. Specifically, financial support should be provided by the home institutions (at the department, college, division level). Participation in broadening participation and informal education activities should be considered in recruiting and promotion decisions, including tenure decisions at universities and other forms of career advancement at non-academic institutions [35]. Finally, institutional leadership should actively and frequently promote and reward such efforts through salary bonuses, fellowships, and awards. This is necessary to avoid placing a disproportionate burden on early-career colleagues and colleagues from marginalized groups.

> **Challenges, Recommendations and Responsibilities**
>
> *Challenge:* Programs aimed at broadening participation in plasma STEM are often mostly focused on teaching technical skills without dedicated attention to building an inclusive community, and often lack an intentional effort to engage with historically excluded communities.
>
> *Challenge:* Such programs are hard to sustain and grow due to inconsistent funding, lack of recognition, and undue burden on early-career colleagues and colleagues from marginalized groups.
>
> *Recommendation:* Broadening participation within the plasma community should be pursued through the establishment and proliferation of ground-up organizations with representative leadership. These organizations should be supported through diverse funding sources and consistent recognition at the institutional and government level.
>
> *Recommendation:* Broadening participation and informal education activities should be adopted as important criteria in recruiting and promotion decisions at the institutional level. Such efforts should be further recognized through bonuses, fellowships, and awards in the home institution.
>
> *Responsibility:* Established and empowered members of the community who assume leadership positions in their institutions should seek to encourage the formation of ground-up organizations with representative leadership, whose work is focused on broadening participations. The leadership should support such organizations financially and formally implement such efforts as criteria for recruitment and promotion.

Another challenge that needs to be addressed by plasma training programs is explaining what a career in plasma can look like, beyond the "traditional" research/teaching academic path. All strategies for broadening participation should implement a clear message on the diverse plasma-relevant career paths, including jobs in industry, national labs, government administration, informal education, etc. In addition, the community should make a dedicated effort to move away from the image that plasma is solely a sub-discipline of physics or engineering. Successful engagement and retention of a diverse plasma workforce would directly benefit from highlighting the cross-disciplinary STEM nature of plasma education and career paths. Insights from research on different groups' perception of STEM careers can be highly beneficial in developing such strategies.

A study by the NSF-funded Innovative Technology Experiences for Students and Teachers (ITEST) project [36] investigated the dispositions and career aspirations related to STEM across diverse groups, including K-12 students and teachers, university students, and STEM professionals. It was established that only STEM professionals and $4^{th}$-$5^{th}$ grade students showed a satisfactory level of positive predispositions toward STEM content areas and STEM as a career. Other groups, including $11^{th}$-$12^{th}$ grade students interested in attending college and college of engineering majors interested in industry jobs, reported significantly less positive dispositions. In addition, all groups exhibited more positive dispositions toward engineering and technology when compared to science and mathematics. Another study on high-school student perception on inclusivity in STEM suggests that STEM is perceived as "pushed" while not necessarily helping, or giving back to their community [37].

Based on discussions among participants at the mini-conference, it was concluded that there has been a historical perception that one needs to have a PhD to work in a plasma-related field, preferably from a renowned plasma program. This faulty perception is harmful to both attracting more students to plasma academic programs and to providing diverse workforce for the national labs and the growing plasma industries. With the increased number of emerging fusion and plasma start-ups [38] and the US endorsing the design of a fusion power plant [39], the demand for employees with various academic backgrounds is also rising. Recruiting quality candidates fast is both a challenge and an opportunity for the plasma community to broaden participation.

Building on findings from STEM research, the plasma community should emphasize the role of plasma as a cross-disciplinary driver of technology by highlighting the various plasma applications in everyday life. Such efforts should start as early as elementary school and should be sustained through all levels of education to establish positive student predispositions towards plasma STEM topics and career paths. These strategies should aim to engage historically underrepresented groups by making plasma a relatable subject, highlighting benefits of plasma education and careers, and bringing opportunities for engagement at the local level. All stakeholder institutions within the plasma community, including academia, national labs, industry, and government should invest in long-term strategic plans and collaborative partnerships focused on broadening participation in the field, offering training and employment opportunities, and establishing a nurturing culture that will ensure success and retention.

One approach is for stakeholder institutions to set specific and measurable goals related to the broadening participation, include these as milestones in their strategic plans, and create dedicated institutional committees tasked with implementing the goals. The plasma community can benefit from the example of other fields who have identified specific goals based on assessment of their workforce needs, such as the Quantum information science and technology community [40], [41]. Another powerful example is the work of the American Institute of Physics (AIP) National Task Force to Elevate African American Representation in Undergraduate Physics & Astronomy (TEAM-UP) [42]. Based on a two-year study of the reasons for the persistent underrepresentation of African Americans in physics and astronomy, TEAM-UP produced a report with its findings and evidence-based recommendations to increase the number of African Americans earning physics and astronomy bachelor's degrees [43]. An example of a bold, actionable, and measurable goal set up by the TEAM-UP report is "to at least double the number of bachelor's degrees in physics and astronomy awarded to African Americans by 2030."

> **Challenges, Recommendations and Responsibilities**
>
> *Challenge:* The breadth of plasma-relevant employment opportunities available across academia, industries, and national labs requires rapid recruitment of a workforce with diverse backgrounds and various levels of academic preparation and subject expertise.
>
> *Recommendation:* Institutional stakeholders within the plasma community should implement into their strategic plan actionable goals with measurable outcomes for broadening participation and form dedicated task force committees to implement these goals.
>
> *Responsibility:* These should be coordinated and collaborative efforts initiated by the leadership in all stakeholder institutions, including academia, national labs, industry, and government.

3. Plasma Terminology in K-12 and The Public Domain

Successful informal science education activities highly benefit from existing knowledge that teachers/students have on the subject, as familiarity assists knowledge retention [44]. Unfortunately, it has been recognized that the word "plasma" is commonly missing from the K-12 curriculum, even though plasma-relevant concepts (such as the Sun and other stars, the states of matter, and sustainable energy sources) are discussed throughout.

One approach to addressing this issue is to advocate incorporating the word "plasma" in the Next Generation Science Standards (NGSS) [11], which have been currently adopted by twenty states in the US and another twenty-four states have adopted standards based on the National Research Council's Frame. Organizations and programs working with the plasma community and with K-12 educators can initiate and manage coordinated efforts on developing plasma-relevant content appropriate for NGSS. One example of such organization is the Contemporary Physics Education Project (CPEP) [45], develop scientifically accurate and visually stimulating materials and activities adaptable to such K-12 science standards. CPEP shares their plasma and fusion materials at the APS-DPP Teacher's Day [46] with hand's-on fusion and plasma activities for the classroom workshop and presentations at the Student Expo [47]. Another example is PhysicsQuest [48] – an initiative by APS Public Engagement [49] that works with scientists and K-12 educators on the development of subject-specific kits for use in the middle school classroom. APS Public Engagement [49] and other relevant branches of professional communities, such as AIP Public Policy [50], and IEEE Government Engagement [51], can advocate adopting plasma STEM terminology to the formal education communities and help advocate the adoption of NGSS by more states. Plasma colleagues can support these efforts at the local level by working with school boards and advocating to state representatives.

The lack of adequate terminology in K-12 education is a detrimental issue that later propagates as a rare usage of the word "plasma" in the public domain. As discussed in the recent white paper *Just Say Plasma* [52], submitted to the Steering Committee of the Decadal Survey for Solar and Space Physics (Heliophysics) 2024-2033, the term "plasma" needs to be formally brought into science press conferences and news reports targeting the public. These issues represent major setbacks for building the future plasma workforce and advocating increased government funding for plasma research, even though numerous current industries rely on basic knowledge of plasma science and engineering. It is the responsibility of all members of the plasma community to promote the use of the term "plasma" when interacting with colleagues from other fields, government representatives, and the press. In addition, consistently writing and endorsing white papers on the subject can help draw attention to the issue.

| **Challenges, Recommendations and Responsibilities** |
|---|
| *Challenge:* K-12 public education curriculum does not include the word "plasma" when discussing plasma-relevant topics. |
| *Recommendation:* The Next Generation Science Standards (NGSS) should incorporate the term "plasma" and terminology relevant to plasma STEM where appropriate throughout the K-12 curriculum. |
| *Responsibility:* Organizations, such as CPEP and PhysicsQuest, should manage adapting plasma terminology to NGSS. Branches within professional societies, such as APS Public Engagement, AIP Public Policy, IEEE Government engagement, should advocate for adopting plasma terminology in NPSS and advocate adopting NPSS in more states. Members of the plasma community should advocate adopting NGSS to local school boards and sate representatives. |
| *Challenge:* The word "plasma" is often missing from scientific press releases and public announcements. |
| *Recommendation:* The term "plasma" and terminology relevant to plasma STEM should be frequently used in the public domain (including press releases, science articles, etc.) |
| *Responsibility:* All members of the plasma community should promote the use of plasma terminology in communications with other colleagues, government, and the press. |

B. Plasma Network for Engagement and Training (PlasmaNET)

Another major goal of the mini-conference was to initiate building a community-based network of people interested in public engagement, education, broadening participation, and workforce development for plasma. We refer to this as PlasmaNET, or Plasma Network for Engagement and Training, following the example set by other networks within our community (such as LaserNet [53] and MagNet [54] and other emerging networks.) Among other goals, the network can provide support to newly hired plasma outreach coordinators through resources, workshops, shared data, etc. It can also help plasma professionals formalize their broader impact work so that they gain recognition for such work from their departments/institutions. PlasmaNET can learn from the APS Joint Network of Informal Physics Education and Research (JNIPER) [55], which is a community of practice for people engaged in our mission of designing, facilitating, or studying informal physics learning activities and programs. PlasmaNET can be viewed as a plasma-focused analog of JNIPER.

| **Challenges, Recommendations and Responsibilities** |
|---|
| *Challenge:* Efforts to engage K-12 and the public, broaden participation, and improve the workforce for the plasma-relevant fields are disjointed. There is little coordinated support for colleagues newly hired as outreach[2] coordinators in the plasma community. |

---

[2] While throughout the document, we prefer to use the word "engagement", here we use "outreach" as it is often times the title of such job positions.

> ***Recommendation:*** Establish the Plasma Network for Engagement and Training (PlasmaNET) as an ecosystem connecting colleagues working in public engagement, education, broadening participation, and workforce development for plasma.
>
> ***Responsibility:*** Various members of the plasma community should work jointly towards the establishment of PlasmaNET. Those include but are not limited to public engagement coordinators, informal education practitioners, physics/STEM education researchers, plasma researchers engaging in broader impact activities, leaders of plasma-related for-profit and nonprofit organizations, K-12 teachers.

Organizing the APS-DPP mini-conference discussed here aimed to increase the visibility of work in public engagement, education and training, DEIA SJ, and other topics related to the plasma workforce. However, as the topics of DPP mini-conferences change each year, there is a need for a separate conference or workshop, which can be organized by an organization such as PNET for the plasma community. Establishing PNET can also coordinate increasing the number of contributed talks and posters for the APS DPP section category *9.00 Science Education, Public Engagement, and DEI*, which typically receives only a handful of submissions each year. These efforts can further be coordinated with the APS Forum of Outreach and Public Engagement. Similarly, the PNET members can work with other professional communities, such as IEEE NPSS (Nuclear and Plasma Sciences Society), to establish a similar session category at more conferences, including the IEEE ICOPS (International Conference on Plasma Science) and APS GEC (Gaseous Electronics Conference).

An important obstacle to increasing interest in such sessions is the lack of financial support for colleagues who are normally not funded to attend plasma professional conferences (such as K-12 teachers and physics education researchers.) Plasma professional communities, including APS DPP and IEEE NPSS, can be instrumental in providing financial support in the form of travel grants, registration fee waivers, and poster awards to boost participation of such colleagues in their conferences. In addition, these organizations can co-sponsor the organization of a separate annual workshop fully dedicated to the topics discussed here. PNET and its institutional members can help raise funds for these activities, coordinate the organization of events, and manage the distribution of travel grants and awards.

There are numerous examples of other conferences that have significant participation in their public engagement and education sessions, showing the growth potential for such sessions at APS DPP, APS GEC, and IEEE ICOPS. The Materials Research Society (MRS) has a topical cluster in the Fall Meeting on "Broader Impacts" to specifically address the NSF requirements, broadly interpreted as workforce development, public engagement, and informal education. The Astronomical Society of the Pacific (ASP), originally a purely scientific organization that faced competition with the American Astronomical Society (AAS), grew its education focus in its sessions to differentiate it from AAS and eventually became almost entirely focused on education, science communication and public engagement. A large portion of AAS meetings is dedicated to education and science communication. Other examples of incorporating public engagement in the activities of professional societies are the AAS Astronomy Ambassadors program [56] and the AAAS Mass Media Science & Engineering Fellowship [57]. APS DPP, APS GEC, and IEEE ICOPS leadership can benefit from interaction with these professional communities.

> **Challenges, Recommendations and Responsibilities**
>
> *Challenge:* The work of informal educators, science communicators, K-12 teachers, science education colleagues, and colleagues involved in engagement, mentoring, and DEIA SJ activities is not well recognized within the plasma community.
>
> *Recommendation:* The APS Committee on Informing the Public has advocated for institutions to consider this work in hiring and promotion decisions [35]. Other plasma-related professional communities, including IEEE, AAS, AAAS, AIAA and others should write and endorse similar statements.
>
> *Recommendation:* The APS Division of Plasma Physics should boost recruitment of abstracts in Section 9.00 on "Science Education, Public Engagement, and DEIA SJ" by offering travel grants and poster awards, and, eventually, by establishing a session of invited topics within these categories.
>
> *Recommendation:* Other professional plasma conferences, such as IEEE ICOPS and APS GEC, should establish a session category dedicated to science education, public engagement, broadening participation and related topics. Participation in these sessions should be boosted by offering travel grants and poster awards.
>
> *Recommendation:* Plasma professional communities, including APS DPP and IEEE NPSS, should sponsor an annual mini-conference or a workshop focused on DEIA SJ, workforce, education, and public engagement (interrelated topics).
>
> *Recommendation:* The professional plasma communities should work with other communities, such as the AAPT (American Association of Physics Teachers) and the AAAS (American Association for the Advancement of Science), to co-sponsor plasma-related sessions at their meetings. APS DPP should sponsor plasma-focused sessions at the APS March meeting.
>
> *Responsibility:* Leadership within the various plasma-related professional communities should initiate and implement these changes.

In addition to creating a repository of resources, organizing workshops, and boosting participation in broader impact sessions at professional conferences, the proposed PNET can initiate and manage large-scale plasma training programs. As many sub-communities organize summer schools for undergraduate interns (e.g., PPPL's Intro Course on Plasma and Fusion for Interns, LTP summer school, HED summer school, etc.), many topical educational materials already have already been developed. However, these training programs are typically geared towards helping the students conduct research and do not grant a formal certificate or degree. The growing need for a qualified plasma workforce will benefit from the establishment of formal plasma training programs that grant professional certificates, similar to a specialization in a Bachelor's degree or a one-year Master's degree. Such programs can be initiated at the state level, for example, through EPSCoR grants, such as the CPU2AL (2017 – 2023) [58] and the FTPP (2022 – 2027) in Alabama. The FTPP is a collaboration of nine universities and one industry focused on developing plasma technology. This project involves several Historically Black Colleges and Universities (HBCUs) and has a strong focus on workforce development. The establishment of similar programs throughout the EPSCoR states will attract students to the participating institutions, supply a qualified workforce for industry partners, and pave the way for forming a coordinated national plasma training program.

For example, in the last five years, CPU2AL organized three different summer research programs for undergraduate students, namely Alabama Plasma Internship program (ALPIP), Alabama Research Experience for Undergraduates (ALREU), and Corporate Internship Program on Plasma Technology Applications (CIPPTA). For most of these students, these experiences were the first time where they get introduced to the field of plasma physics research and most of them have later presented their work in APS-DPP or have continued in higher studies in the STEM field. In similar lines, the next five-year Alabama EPSCoR program FTPP, plans to continue and expand on these positive experiences of CPU2AL.

> **Challenges, Recommendations and Responsibilities**
>
> *Challenge:* Most plasma training programs for high-school and undergraduate students, such as summer schools, do not grant professional certifications.
>
> *Recommendation:* Programs focused on training the plasma workforce should be established both at the state and the national level. Those should be implemented as certified professional development programs, specializations in Bachelor's degrees, or one-year Master's degrees.
>
> *Responsibility:* Plasma colleagues at academic institutions should work with national labs and industries to develop curriculum and training materials. University leadership can work towards providing certification of these programs. A public-private partnership between government funding agencies and industries can be used to sponsor these programs. Nonprofit plasma organizations and emerging community networks can help advertise and recruit participants for the programs.

C. Post-Conference Survey

The mini-conference attendees were surveyed to capture their experience level and needs as a proxy for the larger plasma community engaged in the topics covered. Based on the responses, it seems most have had some formal training in public engagement or education. Nearly all job titles expected, given the target audience of the mini-conference, were represented in our survey respondents: University Faculty or Lecturer, Research Staff/Staff Scientist, Education/Outreach/Public Engagement Staff, Formal K-12 Educator, Informal Educator (museum staff, after-school programs, etc.), Communications Staff, and Graduate Student. The only categories not represented were: Administrative Staff, Post-doctoral Fellow, and Undergraduate Student. The survey was anonymous and invited speakers may have been overrepresented in the survey respondents, reflecting our choice of speakers, rather than the larger audience, though all the speakers are representative of the audience targeted.

The survey responses indicated that funding is a major concern. Respondents echoed the discussions held during the mini-conference in calling for recognition and training, especially in terms of "training for how to create a safe learning environment for marginalized students." In addition to calling for professional recognition for this work, responses indicated the need for plenary sessions or invited talks on public engagement and DEIA SJ topics and how "refreshing" it was to "see this type of event at conferences." Respondents also highlighted the need to push each other to always consider equitable access and strive to remove barriers that may uphold systems of oppression. The networking opportunity was identified as a valuable component of the event, and it is our hope that these opportunities will become more frequent and well attended.

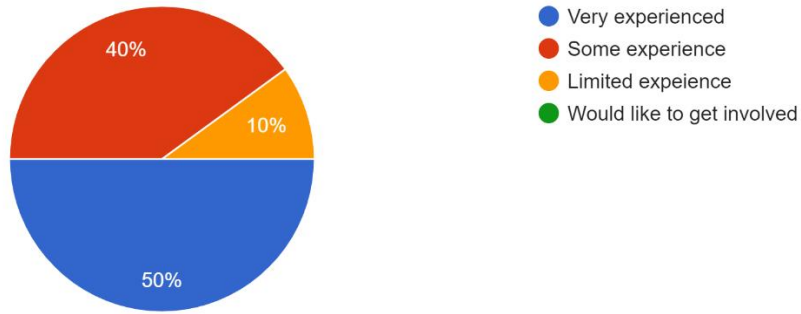

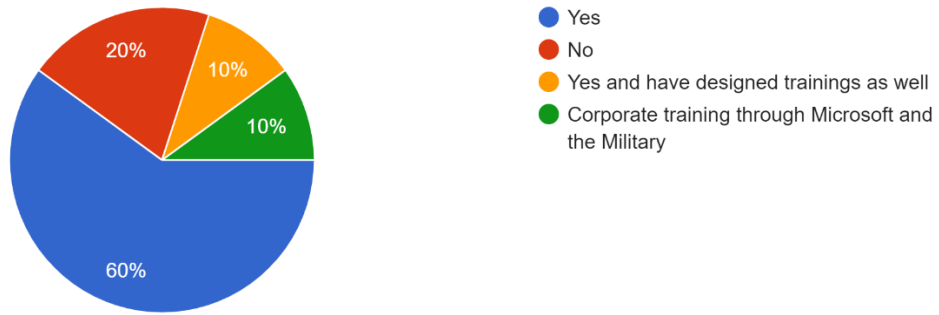

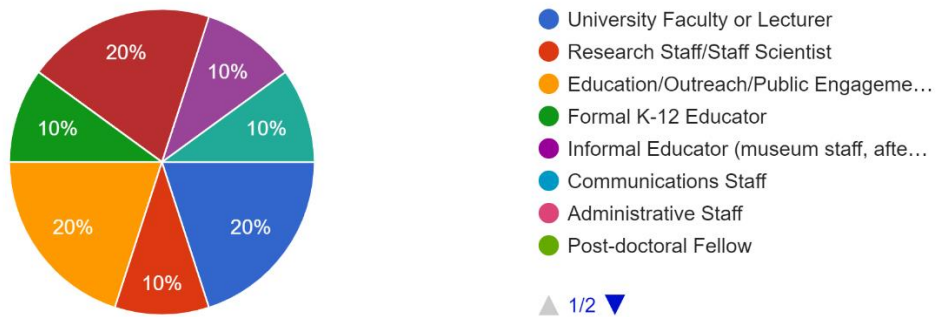

Fig. 3 Results from post-conference survey. a) Participants experience level in public engagement, training/education, DEI, and related topics. b) Participants were asked if they received training or resources for their work in public engagement, training/education, DEI, and related topics. c) Current occupation of participants.

## III. SPECIFIC PLASMA ACTIVITIES

This section provides a list of organizations and initiatives presented at the mini-conference and during related discussions, along with specific projects and more resources. The authors recognize that this list is not exhaustive.

### A. K-12 Informal Plasma Education and Public Engagement

**ISEA.** Informal science education activities (ISEA) have long played an essential role in helping create and recruit the next generation of scientists. Researchers not having constraints from the curriculum, textbooks, and tests have a freedom that allows creative and culturally relevant learning experiences of science. This freedom can present opportunities for engagement by presenting key concepts to the K-12 teachers and students and communicating exciting results from the researcher's work allowing K-12 teachers and students to have more access to science. The audience is the focus of this presentation, which is one of the four stakeholders that benefit from ISEAs – the audiences, the researchers, the institutions, and the field of plasma physics. Colleagues, like Maajida Murdock, who participate in various programs focused on work with K-12 teachers, currently work with the plasma community to promote the use of the two-way interactive method to bring science "alive," spark interest and stimulate curiosity.

**NGSS.** K-12 science courses typically address the Sun and stars, the states of matter, and energy sustainability. Yet, while the Next Generation Science Standards (NGSS) [11] refer to fusion, plasmas are not explicitly mentioned. NGSS are K–12 science content standards, based on the NRC's K-12 Framework for Science Education [6], that set the expectations for what students should know and be able to do. NGSS have been adopted by twenty states and have been developed in collaboration with critical partners, including the National Research Council, the National Science Teachers Association, and the American Association for the Advancement of Science. Another twenty-four states have adopted standards based on the National Research Council's Frame.

**CPEP.** The mission of the non-profit Contemporary Physics Education Project (CPEP) membership of active scientists and practicing teachers is to convey current knowledge of physics topics using scientifically accurate, visually stimulating materials and activities. CPEP's focus topics are particle, cosmological, gravitational, nuclear, and plasma/fusion physics. CPEP shares their Plasma and Fusion materials at the APS-DPP Teachers' Day with a Hands-On Fusion and Plasma Activities for the Classroom workshop and presentations at the Student Expo.

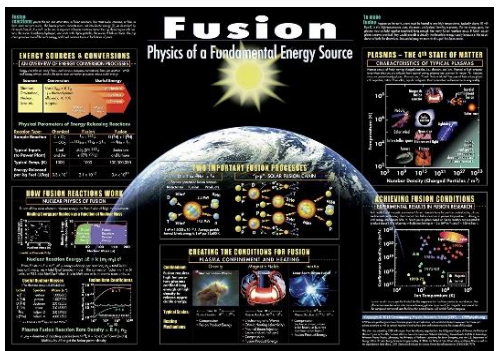

Fig. 4 CPEP Plasma & Fusion chart.

The CPEP Plasma & Fusion chart summarizes topics that are related to plasma and fusion science and serves as a useful classroom teaching device. Development of the chart began in the early 1990's and was supported by the AIP journal, Physics of Plasmas, APS DPP, General Atomics, Lawrence Livermore National Laboratory, MIT, PPPL, University of Rochester Laboratory for Laser Energetics and the U.S. Department of Energy, Office of Fusion Energy Sciences. Individuals from many of these institutions collaborated with CPEP members and other scientists and educators on the original design and content of the chart. The chart is periodically updated and is available in several languages, including French, German, Italian, Spanish and Portuguese. Through contacts made at the 2022 mini-conference discussed in the present paper, plans are underway to translate the chart into Russian, Bulgarian and Hindi.

CPEP members have also developed a collection of engaging plasma and fusion related educational activities. All the necessary instructions, worksheets for students, and teacher's notes for the activities, are accessible for free on their website. Activities explore such things as the motion of charged particles in fields, the electric field near plasma globes, and the properties of plasma through the study half-coated fluorescent bulbs. The CPEP fusion simulation model uses Velcro and bottle-caps and allows students to explore the parameters that influence fusion reactions. While most activities are designed for high school students and introductory undergraduates, there are several activities for middle school students. In addition, on their website they maintain a Resource page with links to plasma-related educational materials.

Since the goal of the group is to convey current knowledge of physics, they are open to collaboration with other engagement groups, institutions, and research groups to help spread knowledge of plasma physics.

**Bryan Stanley's Research.** Informal physics programs create learning environments where physicists, physics students, and public audiences can interact with each other. The formats of these spaces vary considerably with examples including summer camps, public lectures, open houses, and demonstration shows. Participation in informal physics spaces can be impactful for both the public audiences and the program facilitators. Bryan Stanley's qualitative research study investigates the career paths of university students who volunteered in informal physics programs. The study hypothesized that participation in informal programs provided impactful experiences that influenced student volunteers' career paths. It involved interviews with alumni who had volunteered in informal physics programs as students about their career pathways and experiences. The data was analyzed for common themes to better understand the commonalities in experiences for volunteers more broadly. The resulting themes could be useful to informal physics practitioners in better understanding and improving their programs to support their student volunteers.

**The Northwest Nuclear Consortium.** The Northwest Nuclear Consortium is an organization based in Washington state which uses a research grade ion collider to teach a class of high school students nuclear engineering principles based on the Department of Energy curriculum. For over twelve years its founder, Mr. Carl Greninger, has pioneered a private nonprofit helping high school students win over $800,000.00 in scholarships through an educational program that looks more like a company than a classroom. The students achieve experiential technical prowess by growing a Farnsworth fusor into a cutting-edge fusion research device with a unique plasma electrode design offering $2 \times 10^6$ neutron output, which is a remarkable achievement in DD fusion within the ranks of amateurs. The device is then leveraged for oncology, material science, and plasma physics investigations. Most recently, the telemetry has seen serious development, including a Langmuir probe, AZURE Power BI, and touch screen reactor control. Yet, it is the PM role-based student-led engagement that drives the educational experience, as the student researcher quickly realizes that science at this level demands skills far beyond a Langmuir probe. They must become accountants, publicists, advocates, stakeholders, managers, and technologists. The value of this skill set becomes obvious in R&D companies where research dollars are hard to come by.

## B. Diversity, Equity, Inclusion, Accessibility, and Social Justice

**CRANE.** The growth of the nuclear fusion industry and programs in the past few years has shown a need for increased support for young people to join this developing workforce. However, the lack of diversity within the plasma physics community has always been detrimental to encouraging underrepresented, marginalized and students of color to join the field of plasma physics. Although the worst of the COVI-19 pandemic is over, the long-term effects (medical and economic) have disproportionately affected marginalized communities and created an even larger barrier to joining this specific field. The Computational Research Access Network (CRANE) was formed to address these issues. CRANE is a group of graduate students, postdocs, and professionals from marginalized communities. CRANE's goals are to introduce underrepresented students to physics (specifically plasma and nuclear) at early stages, teach the computational tools needed to succeed in both computational and experimental research, and provide them with opportunities in the form of internships or undergraduate research positions. In addition, CRANE's diverse group members work at removing entry barriers for plasma and nuclear physics such as: representation, role models, mentorship, and encouraging culture.

**Franklin Dollar's Program.** Despite many longstanding efforts, representation of numerous demographics in plasma physics in particular and physical sciences as a whole lags behind higher education in general. While many programs focus on the development of the student to address perceived gaps in ability and performance, we instead developed programs in which the primary focus is establishing community and relevance to the student. Franklin Dollar's program explores a variety of modes of student engagement, including near peer and peer mentorship, culturally relevant research experiences, and workshops enabling personal growth and wellness.

**Earl Scime's Programs.** Research studies of persistence in STEM after youth engagement in STEM activities have demonstrated that long-term, sustained programs have the most significant impact on future persistence in STEM. For example, compared to a baseline peer group with the same socio-economic and educational backgrounds, students who participate in a FIRST Robotics program are 50% more likely to attend college, twice as likely to major in science or engineering in college, and four times as likely to end up in an engineering career. Therefore, Earl Scime's focus in rural West Virginia has been to develop long-term, sustainable, hands-on, engineering focused STEM programs through schools, scouting organizations, 4-H, and direct parental engagement. Since 2008, 100% of our local (in their county and three surrounding counties) program graduates have gone on to postsecondary education (98% to college), nearly all full or partial tuition scholarship. 83% have gone on to undergraduate programs in science or engineering. Of those that have gone on to undergraduate programs in science or engineering, over 40% are female or from other underrepresented groups in STEM careers.

## C. User Facilities, Collaborations, and Networks

**SCPC.** Establishing research programs at smaller schools offers the plasma community a chance to increase exposure to plasma science throughout the country, particularly to a diverse undergraduate population. Plasma colleagues in small college can share valuable personal experiences in joining and starting an experimental plasma research program at primarily-undergraduate institutions. The Small College Plasma Consortium (SCPC) is an organization whose aims are to create a community for existing and future plasma researchers at small schools, in addition to providing a platform for shared resources and collaboration amongst small schools and with larger academic institutions. The SCPC has developed a set of recommendations for what the plasma community and funding agencies can do to support the growth and development of these unique research programs.

**ZEUS.** In alignment with recommendations from the National Academy of Science study on high-intensity lasers [59], NSF has recently invested in three extreme light mid-scale research facilities, including the Zettawatt-Equivalent Ultrashort Pulse Laser System (ZEUS) at the University of Michigan. Critical to the success of these facilities is the ability to integrate their research into their outreach efforts. They must develop outreach programs that raise awareness, effectively communicate to various audiences, and promote meaningful understanding of the significant impacts of extreme light research on society. ZEUS's outreach coordinator, Elizabeth Oxford, has established an outreach program at the ZEUS facility, including activities, evaluation methods, and resources. Colleagues like Elizabeth Oxford can share valuable experiences as new members of the plasma-focused outreach community and suggest effective methods for sharing best practices. Those include a virtual toolkit, which would provide members with examples of successful programs, assessments and corresponding data, a directory of members and their areas of expertise, webinars, and other resources.

**LaserNet.** LaserNetUS, a network of 10 high-power laser facilities across America operating effectively as a user facility, was established by the U.S. Department of Energy in 2018. The vision of LaserNetUS is to advance the frontiers of ultra-intense laser science and their multi- and interdisciplinary applications in various sectors including high energy density, materials, and biomedical science. The mission of LaserNetUS is to re-establish US scientific competitiveness in high-energy-density and high-field optical science by advancing the frontiers of laser-science research, providing students and scientists with broad access to unique facilities and enabling technologies, and fostering collaboration among researchers and networks from around the world. Now in its 4th experimental cycle, LaserNetUS has welcomed more than 400 single users, with a quarter being students. Networks like LaserNetUS can share their valuable experience in promoting collaborations and workforce development in the field of plasma, high energy density, and laser science.

**FTPP.** A 2021 study from the National Science Board finds that the performance of U.S. elementary and secondary students in STEM education continues to lag behind that of students from other countries [60]. FTPP (Future Technologies and Enabling Plasma Processes) and the CPU2AL (Connecting the Plasma Universe to Plasma Technology in Alabama) are two projects funded by the NSF Established Program to Stimulate Competitive Research (EPSCoR) that have focused substantial effort in addressing this problem in Alabama. Outreach coordinators for these programs, such as the FTPP outreach coordinator, Laura Provenzani, can provide valuable perspectives on the critical areas that contribute to the success of this effort.

In state-wide projects, such as the EPSCoR programs, the management of the organizational activities requires hiring specialized personnel. The turnover rate of the supporting staff can negatively affect the specialization level and the team's effectiveness. Moreover, building an extended network of external contacts with Agencies or other Universities is essential to success. Universities are not all the same – they vary by size, cultural variation, and student population. In addition, the local environment must also be considered to obtain and maintain cooperation with industries, museums, and non-profit organizations. These activities must be meticulously planned, organized, and monitored in a practical project management framework. One of the main goals is to build additional capacity by creating new educational and outreach programs for college and K-12 students. Consequently, increasing public awareness and plasma-focused education in a state, like Alabama, is challenging. This situation requires action directed towards elementary and secondary students as soon as possible, considering how they vary by size, cultural background, and underrepresented minorities. The FTPP outreach team learned that "marketing" campaigns must be focused and specifically built toward a variegated target. Response rates can differ depending on the channel and methodology used to increase student engagement.

**JNIPER.** Within physics, most of us have, in some form or another, facilitated public engagement, outreach, and/or science communication. However, research on informal physics education (IPER) is relatively nascent. The Joint Network for Informal Physics Education and Research (JNIPER) is a response to a need for articulating and discussing a number of issues in informal physics education. The network will bring together physicists who facilitate informal physics learning activities, along with researchers who investigate the impact of these activities, to align and centralize the informal learning efforts of the physics community at large. Therefore, a big component of this network would focus on building capacity for research and practices to bridge the gap between researchers and practitioners.

**Shams El-Adawy's Research.** The pathways and engagement of physicists in informal physics education are varied, making their professional development needs not well understood. As part of ongoing efforts to build and support the community in the informal physics space, we conducted interviews with physics practitioners and researchers with a range of different experiences. Through thematic analysis, we use personas, a user-centered design tool that stems from the interface design context, to articulate the needs and pain points of professional physicists. Shams El-Adawy's research focuses on a set of four personas: the physicist who engages in informal physics for self-reflection, the physicist who wants to spark interest in physics, the physicist who wants to provide diverse role models to younger students and inspire them to pursue a STEM career, and the physicist who wants to improve the relationship between scientists and the public. This work can help inform tailored professional development resources for building an inclusive network of practitioners in informal science education and public engagement.

## ACKNOWLEDGEMENTS

The mini-conference was co-sponsored by Coalition for Plasma Science (CPS) and the Princeton Plasma Physics Lab. We would like to recognize help by the APS DPP leadership and the contributions of many colleagues who attended the conference and/or participated in the follow-up discussions.